\documentclass[journal]{IEEEtran}
\usepackage[cmex10]{amsmath}
\usepackage{amssymb}
\usepackage{graphicx}
\usepackage{caption}
\usepackage{cite}
\usepackage{subeqnarray}
\usepackage{diagbox}
\usepackage{booktabs}
\usepackage{multirow}
\usepackage{multicol}
\usepackage{algorithm}
\usepackage{algorithmic, setspace}
\usepackage{url}
\usepackage{xcolor}
\usepackage{array}
\usepackage{footnote}
\usepackage{enumerate}
\usepackage{framed}
\usepackage{lineno}
\usepackage{threeparttable}
\usepackage{geometry}
\makesavenoteenv{tabular}

\newcommand{\mb}{\mathbf}
\newcommand{\mc}{\mathcal}

\begin{document}
	\title{Memory Efficient Mutual Information-Maximizing Quantized Min-Sum Decoding for Rate-Compatible LDPC Codes\\
		\text{\Large(Extended Version)}
}
	\author{
		\IEEEauthorblockN{Peng Kang, Kui Cai, Xuan He, and Jinhong Yuan}
	\vspace{-0.6cm}}
	\maketitle
	\begin{abstract}
		In this letter, we propose a two-stage design method to construct memory efficient mutual information-maximizing quantized min-sum (MIM-QMS) decoder for rate-compatible low-density parity-check (LDPC) codes.
		We first develop a modified density evolution to design a unique set of lookup tables (LUTs) that can be used for rate-compatible LDPC codes.
		The constructed LUTs are optimized based on their discrepancy values and a merge function to reduce the memory requirement.
		Numerical results show that the proposed rate-compatible MIM-QMS decoder can reduce the memory requirement for decoding by up to $94.92 \%$ compared to the benchmark rate-compatible LUT-based decoder with generally faster convergence speed.
		In addition, the proposed decoder can approach the performance of the floating-pointing belief propagation decoder within $0.15$ dB.
	\end{abstract}
	%
	\begin{IEEEkeywords}
		Low-density parity-check (LDPC) codes, lookup table (LUT), mutual information-maximizing (MIM), rate-compatible.\vspace{-0.3cm}
	\end{IEEEkeywords}

	\IEEEpeerreviewmaketitle


	\section{Introduction}
	Low-density parity-check (LDPC) codes \cite{Gallager62} have been widely applied to many practical applications, such as wireless communication and data storage systems \cite{Divsalar2009protograph,Nguyen2012protograph,chen2018rate}, for their powerful error correction capability under iterative message passing decoding \cite{Richardson01design}.
	Many decoding algorithms have been developed for the LDPC codes to achieve a good trade-off between the performance and decoding complexity \cite{Jinghu2005reduced,Kang2020eqml}.

	Recently, lookup table (LUT)-based decoders \cite{Nguyen2018faid, Romero2016mim, Meidlinger2020irr, Stark2020ib, wang2020rcq, he2019mutual, kang2020generalized} drew considerable attention due to their simple table lookup operations and excellent error performance with coarse quantization.
	Specifically, the LUT-based decoder in \cite{Nguyen2018faid} is designed based on the density evolution (DE) \cite{Richardson01design}, which aims at minimizing the asymptotic error probability.
	As an alternative, the LUT-based decoders in \cite{Romero2016mim, Meidlinger2020irr, Stark2020ib, wang2020rcq, he2019mutual, kang2020generalized} focus on maximizing the mutual information between the coded bits and the extrinsic messages. 
	In \cite{Romero2016mim, Meidlinger2020irr, Stark2020ib}, several sets of LUTs need to be constructed respectively to update the nodes with different degrees at each iteration.
	This causes a large memory requirement for storing the LUTs when the node degree or the decoding iteration becomes large.
	To reduce the memory usage, the mutual-information-maximizing quantized min-sum (MIM-QMS) decoder was proposed in \cite{kang2020generalized}, which utilizes the integer additions and one set of LUTs to update all nodes with different degrees.

	On the other hand, many practical communication and data storage systems prefer to adopt the rate-compatible codes instead of the fixed-rate codes, e.g., \cite{Nguyen2012protograph,chen2018rate,zhen2018polar}.
	This is because the rate-compatible LDPC codes can improve the system reliability more efficiently by adapting different code rates according to the channel conditions \cite{Nguyen2012protograph}.
	However, most of the LUT-based decoders e.g., \cite{Meidlinger2020irr} and \cite{kang2020generalized}, are designed for an LDPC code with a particular code rate, which is not capable of supporting rate-compatibility.
	Although the rate-compatible LUT-based decoder in \cite{Stark2020ib} is able to reuse the LUTs of several LDPC codes across different code rates, it suffers from performance loss, which increases with the number of reused LUTs due to the mismatch problem.
	Moreover, there are a large number of intermediate LUTs required for the decoding process in \cite{Romero2016mim, Meidlinger2020irr, Stark2020ib}, which vary with both node degrees and iterations.
	Obviously, this increases the memory requirement for hardware implementation.

	In this letter, we aim at designing a memory efficient MIM-QMS decoder for the rate-compatible LDPC codes.
	To this end, we propose a two-stage design method for the construction of the LUTs.
	In particular, we modify the MIM-DE presented in \cite{kang2020generalized} by considering the joint degree distributions of all target LDPC codes rather than their individual degree distributions.
	Compared to the LUTs designed in \cite{Meidlinger2020irr} and \cite{kang2020generalized}, this allows us to construct a unique set of iteration-varying LUTs used for various code rates.
	We further propose a metric, named as the discrepancy value, and a merge function to optimize the constructed LUTs with similar entries.
	Instead of minimizing the degradation of asymptotic performance for reducing the memory variation as in \cite{Meidlinger2020irr}, our proposed LUT optimization method focuses on maximizing the mutual information between the coded bit and the extrinsic messages. 
	We demonstrate that the proposed MIM-QMS decoder can save the memory requirement by up to $94.92 \%$ compared to the rate-compatible LUT-based decoder \cite{Stark2020ib} with faster convergence speed in the moderate-to-high signal-to-noise (SNR) region.
	Moreover, it shows only minor performance degradation compared to the MIM-QMS decoders designed for fixed-rate codes, and also approaches the performance of the floating-pointing belief propagation (BP) decoder.
	\section{Preliminaries}
	\label{section: preliminaries}
	\subsection{Notations}
	In this letter, calligraphy capitals are used to define an alphabet set.
	Normal capitals denote the random variables.
	Lower-case letters denote the realization of a random variable.
	Boldface letters are used to define a vector.\vspace{-0.3cm}
	\subsection{The MIM-QMS decoder}
	Consider a ($q_m$, $q_v$) MIM-QMS decoder \cite{kang2020generalized}, where $q_m$ is the precision of the exchanged messages between check nodes (CNs) and variable nodes (VNs), and $q_v$ is the precision of a posteriori probability (APP) messages. 
	Note that we set $q_v > q_m$ to avoid clipping because the APP messages are generally larger than the exchanged messages.
	During the decoding iterations, the MIM-QMS decoder processes the messages belonging to certain alphabet sets.  
	Define $\mc{L} = \{0, 1, \ldots, |\mc{L}|-1\}$ as the alphabet set of the channel output, where $|\cdot|$ is the cardinality of the set.
	Denote the alphabets of variable-to-check (V2C) and check-to-variable (C2V) messages by $\mc{R} = \{0, 1, \ldots, |\mc{R}|-1\}$ and $\mc{S} = \{0, 1, \ldots, |\mc{S}|-1\}$, respectively.
	In this letter, we consider $|\mc{R}|=|\mc{S}|=|\mc{L}|=2^{q_m}$.
	As depicted in Fig. \ref{fig: general_struct_QMS}, the MIM-QMS decoder performs reconstruction, calculation, and quantization steps for decoding, which can be described correspondingly by three types of functions:
	\begin{figure}[t!]
		\centering
		\includegraphics[width=2.1in]{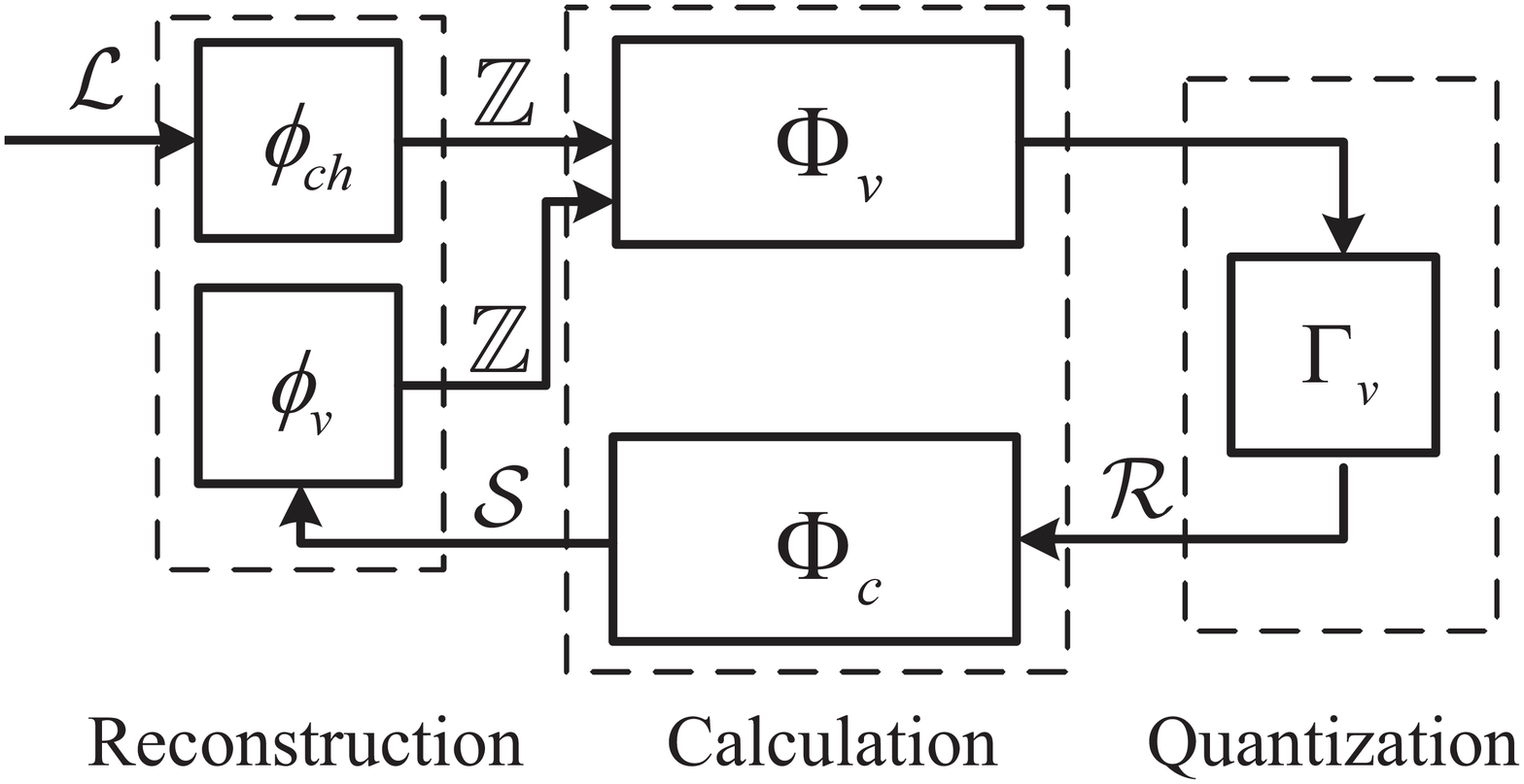}
		\vspace{-0.2cm}
		\caption{The framework of the MIM-QMS decoder.}
		\vspace{-0.7cm}
		\label{fig: general_struct_QMS}
	\end{figure}
	\subsubsection{Reconstruction}
	The reconstruction function maps a $q_m$-bit message to a specific number in the integer domain $\mathbb{Z}$.  
	We denote the reconstruction functions of the channel output and the C2V message by $\phi_{ch}$ and $\phi_v$, respectively.
	\subsubsection{Calculation}
	The computing function for the CN update takes the $q_m$-bit V2C messages as the input, and outputs the $q_m$-bit C2V messages based on the cluster MS algorithm \cite{Meidlinger2020irr}.
	For the VN update, the computing function takes the $q_m$-bit channel output with all other $q_m$-bit reconstructed C2V messages to calculate the V2C messages with $q_v$-bit precision.
	\subsubsection{Quantization}
	The quantization function quantizes a $q_v$-bit V2C message into a $q_m$-bit message based on a threshold set denoting by $\Gamma_v$.
	Here the threshold set is determined by the dynamic programming \cite{he2021dynamic} aiming to maximize the mutual information between the coded bit and the V2C message.

	Given a discretized binary-input additive white Gaussian noise channel (AWGNC) with the noise standard deviation $\sigma_d$ and the degree distributions of an LDPC code with a specific code rate, the MIM density evolution (MIM-DE) is used in \cite{kang2020generalized} to construct the LUTs associated to the reconstruction functions and the threshold set by tracking the evolution of the probability mass functions (pmfs) at each iteration.  
	Note that $\sigma_d$ is known as the design noise standard deviation, which is carefully selected to maximize the mutual information between the coded bit and the V2C message for a preset maximum decoding iteration.\vspace{-0.3cm}
	\section{Design of memory efficient MIM-QMS decoders for Rate-compatible LDPC codes}
	In \cite{Romero2016mim, Meidlinger2020irr, wang2020rcq} and \cite{kang2020generalized}, the LUT-based decoders and the MIM-QMS decoders are designed for an LDPC code with a fixed code rate.
	We refer to these decoders as the rate-specific decoders.
	When different LDPC codes are adopted by the system, these rate-specific decoders such as \cite{Meidlinger2020irr} and \cite{kang2020generalized} need to store multiple sets of LUTs corresponding to different LDPC codes for decoding.
	This increases the memory demand for hardware implementation.
	In this section, we propose a two-stage design method to construct a unique set of LUTs for the MIM-QMS decoder, which can be used for decoding rate-compatible LDPC codes with significantly reduced memory requirement.\vspace{-0.3cm}
	\subsection{Stage 1: LUT Design Based on the Modified MIM-DE}
	In \cite{kang2020generalized}, the MIM-QMS decoders are designed by using the MIM-DE based on the degree distributions of one target LDPC code.
	To construct the MIM-QMS decoder for rate-compatible LDPC codes, we first propose a modified MIM-DE, which considers the degree distributions of all target LDPC codes simultaneously.
	Denoting the sets of the CN and VN degrees of $K$ target LDPC codes by $\mc{\tilde D}_c$ and $\mc{\tilde D}_v$, we have \vspace{-0.05cm}
	\begin{equation}
	\begin{aligned}
		\mc{\tilde D}_c &= \bigcup\nolimits_{k = 1}^K {{\mc{D}_{c,k}}} =\{d_{c, 1},  d_{c, 2}, \ldots, d_{c, \text{max}}\} , \\ \mc{\tilde D}_v &= \bigcup\nolimits_{k = 1}^K {{\mc{D}_{v,k}}}=\{d_{v, 1}, d_{v, 2}, \ldots, d_{v, \text{max}}\}
	\end{aligned}\vspace{-0.05cm}
	\end{equation}
	where $\mc{D}_{c, k}$ and $\mc{D}_{v,k}$ refer to the sets of the CN and VN degrees of the $k$-th LDPC code, respectively.
	Define the joint degree distributions as \vspace{-0.05cm}
	\begin{equation}
	\label{degreeDis}
	{\tilde \rho}(\zeta) = \sum_{i \in \mc{\tilde D}_c} {\tilde \rho}_i \zeta^{i-1}, \, {\tilde \theta}(\zeta) = \sum_{j \in \mc{\tilde D}_v} {\tilde \theta}_j \zeta^{j-1},\vspace{-0.05cm}
	\end{equation}
	where ${\tilde \rho}_i$ and ${\tilde \theta}_j$ are the fractions of edges incident to the CNs with degree-$i$ and the VNs with degree-$j$, respectively, by considering the Tanner graphs \cite{Tanner81} of all target LDPC codes.
	We denote the random variable for the coded bit by $X$, which takes values from $\mc{X} = \{0, 1\}$.
	Denote the random variable for the channel output by $L$, and the random variable for the V2C (resp. C2V) message by $R$ (resp. $S$).
	Define $P_{L|X}$ as the pmf of channel output $L$ conditioned on the coded bit $X$.
	Define $P_{R|X}$ and $P_{S|X}$ as the pmfs of the V2C message $R$ and the C2V message $S$ conditioned on the coded bit $X$, respectively.
	\begin{figure}[t!]
		\centering
		\includegraphics[width=2.18in]{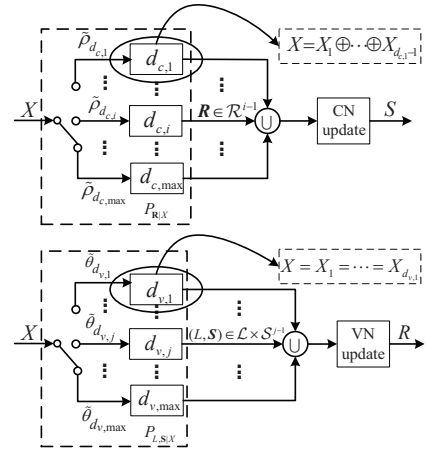}
		\vspace{-0.1cm}
		\caption{The framework of the MIM-QMS decoder for decoding the rate-compatible LDPC codes.}
		\label{QMS_update}
		\vspace{-0.7cm}
	\end{figure}

	As illustrated by Fig. \ref{QMS_update}, for the CN update, the V2C messages are classified into $|\mc{\tilde D}_c|$ alphabet sets correspondingly to different CN degrees.
	The modified MIM-DE aims to compute the conditional pmf $P_{S|X}$ based on the joint CN degree distribution. 
	Denote the vector of V2C messages sent to a degree-$i$ ($i \in \mc{\tilde D}_c$) CN by $\mb{R} \in \mc{R}^{i-1}$.
	Define $\mb{r}=(r_1,r_2,\ldots,r_{i-1}) \in \mc{R}^{i-1}$ as a realization of $\mb{R}$.
	According to the independent and identical distribution (i.i.d) assumption \cite{Richardson01design}, the joint pmf $P_{\mb{R}|X}$ at a degree-$i$ CN is given by \vspace{-0.1cm}
	\begin{equation}
	\label{eqn: joint P_R|X}
	P_{\mb{R}|X}(\mb{r}|x) = \left(\frac{1}{2}\right)^{i - 2} \sum_{\mb{x}: \oplus \mb{x} = x} \prod_{k = 1}^{i-1} P_{R|X}(r_k|x_k),
	\vspace{-0.1cm}
	\end{equation}
	where $x \in \mc{X}$ is a realization of $X$, and $\mb{x} = {\rm{ }}({x_1},{x_2}, \ldots ,{x_{i - 1}})$ is the vector of coded bits associated to the neighboring VNs.
	Here $\oplus \mb{x} = x$ means the checksum of the CN is satisfied.
	Denote a realization of the C2V message $S \in \mc{S}$ by $s$.
	For each incoming realization $\mb{r}$, the computing function $\Phi _c$ calculates the corresponding output $s$ by performing the MS operation \cite{Meidlinger2020irr}.
	Considering all possible values $s \in \mc{S}$ for each CN alphabet set and the fraction ${\tilde \rho}_i$ in (\ref{degreeDis}), the V2C messages in each CN alphabet set contribute to the  output C2V messages with probability ${\tilde \rho}_i$.
	Therefore, we obtain
	\begin{equation}
	\label{psx_ms}
	{P_{S|X}}(s|x) = \sum\limits_{i \in {\mc{\tilde D}_c}} {{\tilde \rho _i} \cdot \!\! \sum\limits_{\mb{r} \in {\mc{R}^{i - 1}}, \hfill\atop \Phi _c(\mb{r}) = s} {{P_{\mb{R}|X}}} (\mb{r}|x)},\vspace{-0.1cm}
	\end{equation}   
	where $\Phi _c$ is given by \cite[Eq. (16)]{kang2020generalized}\vspace{-0.1cm}
	\begin{equation}
	\label{eqn: def of Phi_c ms}
	{\Phi _c}(\mb{r}) = f^{-1}\left( {\prod\limits_{k = 1}^{{i} - 1} sgn(f({r_k}))} \! \cdot \!\!\!\! \mathop {\min }\limits_{k = 1,2, \ldots, {i} - 1} (\left| {f({r_k})} \right|)\right).\vspace{-0.1cm}
	\end{equation}  
	Here $f(\cdot)$ maps the V2C messages in $\mc{R} = \{0, 1, \ldots, {{2^{{q_m}}} - 1}\}$ correspondingly to an integer set $\{ {2^{{q_m} - 1}}, \ldots ,1, - 1, \ldots  - {2^{{q_m} - 1}}\}$, and $f^{-1}(\cdot)$ operates as the inverse function of $f(\cdot)$.

	For the VN update, as shown by Fig. \ref{QMS_update}, the C2V messages are characterized by $|\mc{\tilde D}_v|$ alphabet sets corresponding to different VN degrees.
	The modified MIM-DE calculates the conditional pmf $P_{R|X}$ based on the joint VN degree distribution.
	Denote the vector of the C2V messages and the channel output at a degree-$j$ ($j \in \mc{\tilde D}_v$) VN by $(L,\mb{S}) \in \mc{L} \times \mc{S}^{j-1}$.
	Define $\mb{s}=(s_1,s_2,\ldots,s_{j-1}) \in \mc{S}^{j-1}$ as a realization of $\mc{S}^{j-1}$.
	The joint pmf $P_{L,\mb{S}|X}$ at a degree-$j$ VN can be computed by \vspace{-0.2cm}
	\begin{equation}
	\label{Plsx}
	P_{L,\mb{S}|X}(l, \mb{s}|x) =  P_{L|X}(l|x) \prod_{k = 1}^{j-1} P_{S|X}(s_k|x).
	\vspace{-0.2cm}
	\end{equation}
	Define $\mc{B}$ as the alphabet set of the $q_v$-bit V2C message calculated by the computing function for the VN update, where $B \in \mc{B}$ is the random variable for the $q_v$-bit V2C message.
	Let $P_{B|X}$ be the pmf of $B$ conditioned on the coded bit $X$.
	According to the fraction ${\tilde \theta}_j$ in (\ref{degreeDis}), the C2V messages $(L,\mb{S})$ in each VN alphabet set contribute to the output V2C messages with probability ${\tilde \theta}_j$ so that we have \vspace{-0.1cm}
	\begin{equation}
	\label{Pbx_all}
	P_{B|X}(b|x) = \sum_{j \in \mc{\tilde D}_v} {{\tilde \theta}_j} \cdot \sum_{(l, \mb{s}) \in \mc{L} \times \mc{S}^{j-1}, \hfill\atop \Phi_v(l, \mb{s}) = b} P_{L, \mb{S} | X}(l, \mb{s} | x)
	\vspace{-0.1cm}
	\end{equation}
	for a realization $b$ of $B$.
	Note that the computing function $\Phi _v$ is given by \cite[Eq. (4)]{kang2020generalized} \vspace{-0.2cm}
	\begin{equation}
	\label{eqn: def of Phi_v}
	\Phi_v(l, \mb{s}) = \phi_{ch}(l) + \sum_{k = 1}^{{j} - 1} \phi_v(s_k).
	\vspace{-0.2cm}
	\end{equation}
	By taking $\mc{B}$ and $P_{B|X}$ as inputs, we perform the dynamic programming (DP) \cite{he2021dynamic} to obtain the threshold set $\Gamma_v$ for the VN update, and the conditional pmf of the V2C message $P_{R|X}$.
	We refer $\textbf{DP}(\cdot)$ to the DP operation, which can be described as\vspace{-0.1cm}
	\begin{equation}
	\label{eqn: P_(R|X) lambda}
	[{P_{R|X}},{\Gamma_v }] = \textbf{DP}(\mc{B},{P_{B|X}}).
	\end{equation}
	In addition, the threshold set for bit decision, denoted by $\Gamma_e$ with $|\Gamma_e|=1$, also needs to be designed for each iteration, which is obtained similarly to $\Gamma_{v}$ by considering $(l, \mb{s}) \! \in \! \mc{L} \! \times \! \mc{S}^{j}$. \vspace{-0.8cm}
	\subsection{Stage 2: LUT Optimization}
	In general, the LUT constructed by the modified MIM-DE is assumed to result in distinct LUTs for different iterations, i.e., the reconstruction functions and the threshold sets vary with iterations.
	Apparently, storing these iteration-varying LUTs increases the memory requirement.
	We notice that the LUTs corresponding to the same reconstruction function or threshold set has similar entries for certain consecutive iterations.
	This is because the pmfs computed by the modified MIM-DE hardly change during these iterations.
	A similar phenomenon is also observed in \cite{Richardson01design} for the conventional DE.
	Motivated by this observation, we propose an LUT optimization method in this section, which reuses one optimized LUT for different consecutive iterations to reduce the memory requirement.

	Consider an LUT for either the reconstruction function or the threshold set at the $t$-th iteration, which can be represented by a vector $\mb{y}^{(t)} = (y_1^{(t)}, y_2^{(t)}, \ldots, y_{N}^{(t)})$.
	Here $N$ is the total number of entries in the LUT, which is determined by the bit precision $q_m$.
	More specifically, we have $N = 2^{q_m}$ for the reconstruction function, and $N = 2^{q_m}-1$ for the threshold set.
	For a preset maximum number of iterations $I_\text{max}$, we define the discrepancy of two LUTs as \vspace{-0.05cm}
	\begin{equation}
	\label{deltay}
	d({\mb{y}^{(t)}},{\mb{y}^{(t')}}) = \sqrt {\sum\nolimits_{n = 1}^N {{{(y_n^{(t)} - y_n^{(t')})}^2}} }, \vspace{-0.05cm}
	\end{equation} 
	where $1 \leq t < t' \leq I_\text{max}$.
	A smaller value of the discrepancy means the two LUTs have corresponding entries similar to each other, and vice versa.  
	For the LUTs associated to the $I_\text{max}$ decoding iterations, we first divide them into $M$ groups.
	Define $\mc{I}_m$ as the set of iteration indices associated to the LUTs in the $m$-th ($m = 1,2,\ldots,M$) group.
	For all LUTs with $t \in \mc{I}_m$, we define a merge function \vspace{-0.1cm}
	\begin{equation}
	\label{lutopt}
	f(\{\mb{y}^{(t)}: t \in {\mc{I}_m}\}) = \left[ {\frac{1}{{\left| {{\mc{I}_m}} \right|}}\sum\limits_{t \in {\mc{I}_m}} {{\!\!\mb{y}^{(t)}}} } \!\right], \vspace{-0.1cm}
	\end{equation}
	which computes the optimized LUT by averaging the vectors $\mb{y}^{(t)}, \forall \, t \in \mc{I}_m$.
	We denote the obtained LUT by $\mb{\tilde y}$ and reuse it for any iteration $t \in \mc{I}_m$.
	\begin{table}[t!]
		\begin{algorithm}[H]
			\small
			\setstretch{1}
			\caption{LUT Optimization Process}
			\label{alg:LUTopt}
			\begin{algorithmic}[1]
				\REQUIRE $Q^*$, $I_\text{max}$, $\mb{y}^{(t)}$ for $t=1,2,\ldots,I_\text{max}$
				\ENSURE Optimized LUTs $\mb{y}^{*(t)}$ for $t=1,2,\ldots,I_\text{max}$
				\STATE{$\Delta {d^*} = 0$, $k = 1$}
				\WHILE {$k \leq I_\text{max}-1$}
				\STATE {Compute $\Delta {d^{(k)}} = d({\mb{y}^{(k)}},{\mb{y}^{(k + 1)}})$ by (10)}
				\IF {$\Delta {d^{(k)}} \leq \Delta {d^{*}}$}
				\STATE{$k = k + 1$, go to step 3}
				\ELSE
				\STATE {Set $M=1$, $\mc{I}_1 = \{1\}$}
				\FOR {$t=1:I_\text{max}-1$}
				\IF {$d({\mb{y}^{(t)}},{\mb{y}^{(t + 1)}}) \leq \Delta {d^{(k)}}$}
				\STATE{$\mc{I}_M = \mc{I}_M  \cup \{t+1\}$}
				\ELSE
				\STATE{$M = M + 1$, $\mc{I}_M = \{t+1\}$}
				\ENDIF
				\ENDFOR
				\STATE{Update $\mb{\tilde y}^{(m)} = f(\{\mb{y}^{(t)}: t \in {\mc{I}_m}\})$ based on (11) for $m = 1,2,\ldots,M$}
				\STATE{Run the modified MIM-DE with $\mb{\tilde y}^{(m)}, m = 1,2,\ldots,M$}
				\STATE{Compute the mutual information $Q^{(k)}$ at $I_\text{max}$}
				\IF {$Q^{(k)} \geq Q^*$ and $\Delta {d^{(k)}} \geq \Delta {d^*}$}
				\STATE{$\Delta {d^*} = \Delta {d^{(k)}}$}
				\STATE{$\mb{y}^{*(t)} = \mb{\tilde y}^{(m)}, \, \forall \, t \in \mc{I}_m$}
				\ENDIF
				\STATE{$k = k + 1$}
				\ENDIF
				\ENDWHILE
			\end{algorithmic}
		\end{algorithm}\vspace{-1cm}
	\end{table}
	Note that $\left[ \mb{x} \right]$ returns the closest integer to each element of the vector $\mb{x}$.
	We denote the LUT optimized for the $m$-th group by $\mb{\tilde y}^{(m)}$, which is reused for all iterations $t \in \mc{I}_m$.
	To arrange the LUTs in the $m$-th group, we conduct greedy search by consecutive iterations based on a discrepancy threshold denoted by $\Delta {d^*}$.
	In this letter, $\Delta {d^*}$ is determined by performing exhaustive search among $I_\text{max} \! - \! 1$ discrepancy values.
	We denote the discrepancy value for the $k$-th search by $\Delta {d^{(k)}} = d({\mb{y}^{(k)}},{\mb{y}^{(k+1)}})$ for $k=1,2,\ldots,I_\text{max} \! - \! 1$.
	More specifically, we first consider $\Delta {d^{(k)}}$ as the discrepancy threshold to update $\mb{\tilde y}^{(m)}$ for the $m$-th group.
	After that, we run the modified MIM-DE with $\mb{\tilde y}^{(m)}$ and compute the mutual information between the coded bit and the V2C message at $I_\text{max}$ denoted by $Q^{(k)}$.
	Given the target mutual information value $Q^{*}$, the optimal discrepancy threshold is selected as the maximum discrepancy value to achieve $Q^{(k)} \geq Q^{*}$, where $Q^{*}$ is set to a value approaching 1. 
	The optimization process is summarized in \textbf{Algorithm \ref{alg:LUTopt}}.\vspace{-0.4cm}
	\subsection{Remarks}
	Different from \cite{Romero2016mim, Meidlinger2020irr, Stark2020ib, wang2020rcq, he2019mutual, kang2020generalized}, the joint degree distribution considered by the modified MIM-DE is not specific for a target LDPC code.
	This causes the performance degradation for the rate-compatible MIM-QMS decoders compared to their rate-specific counterparts.
	However, the performance loss can be reduced by carefully selecting the target LDPC codes in the LUT design.
	We quantify $\sigma_d$ equivalently by the corresponding design signal-to-noise ratio (SNR) $\tau  \buildrel \Delta \over =  - 20{\log _{10}}(\sqrt {2{R_c} \cdot \sigma _d^2} )$, where $R_c$ is the code rate of the LDPC code. 
	According to extensive simulation results, the difference of $\tau$ between any two target LDPC codes is limited up to $1.2$ dB in this letter for a desirable performance.
	It is an open problem that whether there exists a systematic way to select the target LDPC codes for acceptable deterioration in performance.  
	\section{Simulation Results}
	\label{section: simulation results}
	In this section, we evaluate the frame error rate (FER) performance of the proposed rate-compatible MIM-QMS decoder via Monte-Carlo simulations. 
	We also investigate the decoding latency aspects and evaluate the memory requirement of the rate-compatible MIM-QMS decoders.
	\begin{table}[h!]
		\LARGE
		\renewcommand{\arraystretch}{1.2}
		\caption{Degree Distributions and Noise Standard Deviation ${\sigma _d}$ for the MIM-QMS Decoders}
		\label{table: simulation parameters}
		\centering
		\resizebox{0.5\textwidth}{!}{
			\begin{tabular}{|c|c|c|c|}
				\hline
				$R_c$        & Degree distributions ($\rho (x), \theta (x)$) & Rate-specific  & Rate-compatible     \\ \hline
				\multirow{2}{*}{$2/3$}        
				& $\rho (x) = {x^{10}}$ & \multirow{2}{*}{$0.7016$} & \multirow{6}{*}{$0.6195$} \\
				& $\theta (x) = {\rm{0}}{\rm{.1591}}x + {\rm{0}}{\rm{.4091}}{x^2} + {\rm{0}}{\rm{.1591}}{x^6} + {\rm{0}}{\rm{.2727}}{x^{7}}$ & & \\ \cline{1-3}
				\multirow{2}{*}{$3/4$}        
				& $\rho (x) = {\rm{0}}{\rm{.3182}}{x^{13}} + {\rm{0}}{\rm{.6818}}{x^{14}}$ & \multirow{2}{*}{$0.6266$} &  \\
				& $\theta (x) = {\rm{0}}{\rm{.1136}}x + {\rm{0}}{\rm{.4091}}{x^2} + {\rm{0}}{\rm{.4773}}{x^5}$ & & \\ \cline{1-3}
				\multirow{2}{*}{$5/6$}        
				& $\rho (x) = {\rm{0}}{\rm{.7412}}{x^{20}} + {\rm{0}}{\rm{.2588}}{x^{21}}$ & \multirow{2}{*}{$0.5494$} & \\
				& $\theta (x) = {\rm{0}}{\rm{.0706}}x + {\rm{0}}{\rm{.1765}}{x^2} + {\rm{0}}{\rm{.7529}}{x^3}$ & & \\ \hline
			\end{tabular}
		}
		\vspace{-0.7cm}
	\end{table}
	\subsection{FER Performance}
	\label{simFER}
	\vspace{-0.1cm}
	Assume that binary LDPC codewords are modulated by binary phase-shift keying (BPSK) and transmitted over the AWGNC.
	We denote the floating-point precision by ``$\infty$''.
	For comparison, we include the FER performance of the ($4$, $8$) rate-specific MIM-QMS decoders \cite{kang2020generalized}, the 4-bit rate-compatible LUT-based decoder in \cite{Stark2020ib}, the 4-bit normalized min-sum (NMS) decoder \cite{Jinghu2005reduced}, and the BP($\infty$) decoder.
	Note that the LUT-based decoder is designed based on the MIM quantization by using the dynamic programming \cite{he2021dynamic} \footnote{Here we consider the dynamic programming instead of the information bottleneck method in \cite{Stark2020ib} to realize the MIM quantization because the dynamic programming is proved to be optimal to maximize the mutual information \cite{he2021dynamic}}, and the scaling factor of the NMS decoder is optimized as $0.75$.
	The degree distributions of the simulated LDPC codes and the design noise standard deviation ${\sigma _d}$ for the associated MIM-QMS decoders are given in TABLE \ref{table: simulation parameters}.
	We set $I_{\text{max}} = 30$ for all decoders.
	\begin{figure}[!t]
		\centering
		\includegraphics[width=2.7in]{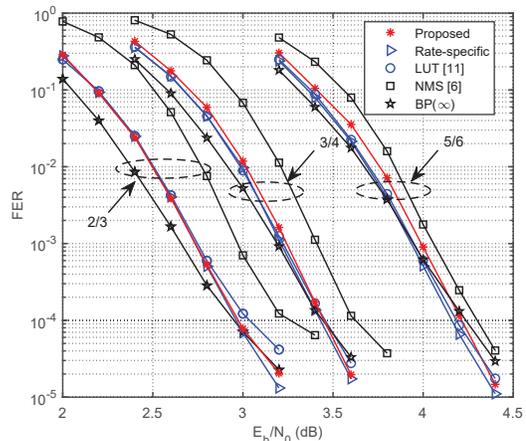}
		\vspace{-0.2cm}
		\caption{FER performance of the length-$1296$ IEEE 802.11n LDPC codes \cite{IEEESTD802_11n} with $R_c$ = $2/3, \, 3/4$, and $5/6$.} 
		\vspace{-0.7cm}
		\label{fig: 802_11n}
	\end{figure}

	Fig. \ref{fig: 802_11n} depicts the FER performance of the length-$1296$ IEEE 802.11n LDPC codes \cite{IEEESTD802_11n} with code rates $R_c$ = $2/3, \, 3/4$, and $5/6$, respectively.
	We can see that our proposed rate-compatible MIM-QMS decoders achieve almost the same FER performance compared to the rate-specific MIM-QMS decoders for each code rate.
	Note that the performance gap between the rate-compatible MIM-QMS decoder and its rate-specific counterparts becomes larger with the increase of the code rates.
	This is because the rate-compatible MIM-QMS decoder is designed based on the joint degree distributions with respect to all target LDPC codes, which considers a larger portion of the VNs with low degrees for higher code rates compared to the rate-specific design.
	These low-degree VNs yield a slower convergence speed.
	Therefore, the mismatched VN degree distribution between the decoding and the LUT design leads to slightly more degradation in performance for higher code rates.
	Moreover, the rate-compatible MIM-QMS decoder achieves the FER performance close to the LUT-based decoder \cite{Stark2020ib}, and even outperforms the LUT-based decoder by about $0.1$ dB in the high SNR region for $R_c$ = $2/3$.
	In addition, the rate-compatible MIM-QMS decoder surpasses the NMS decoder by up to $0.3$ dB, and approaches the performance of the BP($\infty$) decoder within $0.15$ dB for all code rates.
	Note that the BP($\infty$) decoder shows worse FER performance than the rate-compatible MIM-QMS decoders in the high SNR region.
	This is due to the existing of the degree-2 VNs in the code graph of the IEEE 802.11n LDPC codes.
	The cycles confined among these degree-2 VNs become low-weight codewords, which are the most detrimental objects for floating-point decoders \cite{Richardson03}.
	\vspace{-0.4cm}
	\subsection{Decoding Latency Analysis}
	\vspace{-0.1cm}
	To evaluate the decoding latency of the proposed rate-compatible MIM-QMS decoder, we first compare the convergence speed of different decoders in Section \ref{simFER}.
	Denote the average number of iterations required for decoding one codeword by $I_\text{avg}$.
	TABLE \ref{table_Iavg} compares $I_\text{avg}$ with respect to $E_b/N_0$.
	We can see that the rate-compatible MIM-QMS decoder requires the least $I_\text{avg}$ for most of the simulated SNRs for each code rate.
	\begin{table}[h!]
		\centering
		\renewcommand{\arraystretch}{1.2}
		\Large
		\caption{Comparison of $I_\text{avg}$ for Different LDPC Decoders}
		\label{table_Iavg}
		\resizebox{0.5\textwidth}{!}{
		\begin{tabular}{|cc|c|c|c|c|c|c|c|}
			\hline
			\multicolumn{2}{|c|}{$E_b/N_0$ (dB)}                                			& 2     & 2.2   & 2.4    & 2.6   & 2.8   & 3     & 3.2  \\ \hline
			\multicolumn{1}{|c|}{\multirow{2}{*}{$I_\text{avg}$}} & Proposed 		&19.2   &13.99  &10.61   &8.36   &6.94   &5.96   &5.24  \\ \cline{2-9} 
			\multicolumn{1}{|c|}{}                      		  & Rate-specific   &19.25  &14.33  &11.23   &9.15   &7.78   &6.79   &6.02  \\ \cline{2-9} 
			\multicolumn{1}{|c|}{($R_c = 2/3$)}              	  & LUT \cite{Stark2020ib}        &19.42  &14.96  &11.62   &9.3    &7.68   &6.69   &5.86   \\ \cline{2-9} 
			\multicolumn{1}{|c|}{}                          	  & NMS \cite{Jinghu2005reduced}        &27.1   &21.88  &16.03   &10.87  &7.93   &6.33   &5.34	\\ \hline
			\multicolumn{2}{|c|}{$E_b/N_0$ (dB)}                                   		& 2.4   & 2.6   & 2.8    & 3     & 3.2   & 3.4   & 3.6 	 \\ \hline
			\multicolumn{1}{|c|}{\multirow{2}{*}{$I_\text{avg}$}} & Proposed 		&20.59  &14.68  &10.63   &7.79   &6.16   &5.16   &4.46   \\ \cline{2-9} 
			\multicolumn{1}{|c|}{}                      		  & Rate-specific   &19.47  &14.32  &10.48   &7.98   &6.48   &5.51   &4.8    \\ \cline{2-9} 
			\multicolumn{1}{|c|}{($R_c = 3/4$)}                   & LUT \cite{Stark2020ib}        &20.28  &14.96  &10.98   &8.35   &6.73   &5.66   &4.89    \\ \cline{2-9} 
			\multicolumn{1}{|c|}{}                          	  & NMS \cite{Jinghu2005reduced}        &26.97  &22.3   &15.7    &10.28  &7.09   &5.42   &4.48	\\ \hline
			\multicolumn{2}{|c|}{$E_b/N_0$ (dB)}                                   		& 3.2   & 3.4   & 3.6    & 3.8   & 4     & 4.2   & 4.4    \\ \hline
			\multicolumn{1}{|c|}{\multirow{2}{*}{$I_\text{avg}$}} & Proposed 		&16.16  &10.46  &7.34    &5.31   &4.22   &3.54   &3.06     \\ \cline{2-9} 
			\multicolumn{1}{|c|}{}                      		  & Rate-specific   &15.02  &9.99   &7.14    &5.4    &4.4    &3.74   &3.26     \\ \cline{2-9}  
			\multicolumn{1}{|c|}{($R_c = 5/6$)}              	  & LUT \cite{Stark2020ib}        &15.48  &10.46  &7.35    &5.5    &4.41   &3.71   &3.2     \\ \cline{2-9} 
			\multicolumn{1}{|c|}{}                          	  & NMS \cite{Jinghu2005reduced}        &20.64  &14.39  &9.12    &5.82   &4.24   &3.42   &2.88	\\ \hline
		\end{tabular}
	}\vspace{-0.2cm}
	\end{table}

	We also investigate the decoding time of the proposed rate-compatible MIM-QMS decoder at each iteration.
	Due to the MS operation adopted for all decoders in Section \ref{simFER}, the decoding time is of $O(\left\lceil {{{\log }_2}({d_{c,{\rm{max}}}})} \right\rceil)$ for the CN update per iteration.
	For the VN update of the NMS decoder \cite{Jinghu2005reduced}, the decoding time is of $O(\left\lceil {{{\log }_2}({d_{v,{\rm{max}}}})} \right\rceil)$  for additions while the rate-compatible LUT-based decoder \cite{Stark2020ib} needs $O({d_{v,{\rm{max}}}} )$ time for table lookup operations due to the concatenating structure of LUTs.
	Compared to the LUT-based decoder, the rate-compatible MIM-QMS decoder adopts integer additions with $O(\left\lceil {{{\log }_2}({d_{v,{\rm{max}}}})} \right\rceil)$, and processes the table lookup operations for the threshold set and reconstruction functions in parallel.
	This results in the decoding time of $O(\left\lceil {{{\log }_2}({d_{v,{\rm{max}}}})} \right\rceil + q_m)$ in total for the VN update per iteration.
	Since $q_m$ is small in general, the rate-compatible MIM-QMS decoder only has slightly increased decoding time compared to the NMS decoder, and achieves much less decoding time with respect to the LUT-based decoder.\vspace{-0.3cm}
	\subsection{Comparison of Memory Requirement}
	To demonstrate the memory efficiency of the proposed rate-compatible MIM-QMS decoders, we compare the overall memory requirement of different quantized decoders in Section \ref{simFER} for the IEEE 802.11n LDPC codes \cite{IEEESTD802_11n} with $R_c$ = $2/3, \, 3/4$, and $5/6$.
	We consider that the decoder implementations are based on digital signal processors or software defined radios such that LUTs are stored in memories \cite{Stark2020ib}.
	Note that we classify the memory into either for arithmetic computations or for storing the LUTs.
	For the CN update, all quantized decoders require memory for storing the two C2V messages obtained by the min-sum operation at each CN.
	For the VN update, both MIM-QMS decoders and the NMS decoder need to save the APP messages computed by additions for each VN.
	Moreover, all MIM-QMS decoders and the rate-compatible LUT-based decoder \cite{Stark2020ib} need memory to store the LUTs.
	Without loss of generality, the memory requirement of one LUT is given by $N \cdot q_l$ bits, where $q_l$ refers to the bit precision of one entry in the LUT.  
	According to \cite{Stark2020ib}, the LUTs of the LUT-based decoder have $q_l = {q_m}$ and there are $d_{v, \text{max}}$ and $|\mc{\tilde D}_v|$ LUTs of size $N = 2^{2q_m} $ in total for the VN update and the bit decision at each iteration, respectively.
	The overall memory requirement for different 4-bit decoders with $I_\text{max} = 30$ is summarized in TABLE \ref{table_memory}.
	We observe that the memory requirement for the proposed rate-compatible MIM-QMS decoder only increases by $8.28 \%$ compared to the NMS decoder.
	More importantly, the rate-compatible MIM-QMS decoder can reduce the memory demand by $45.37 \%$ and $94.92 \%$ compared to its rate-specific counterparts and the LUT-based decoder, respectively.\vspace{-0.2cm}
	\begin{table}[h!]
		\centering
		\renewcommand{\arraystretch}{1.2}
		\Large
		\caption{The Overall Memory Requirement for Different 4-bit Decoders with $I_\text{max} = 30$}
		\label{table_memory}
		\resizebox{0.5\textwidth}{!}{
			\begin{tabular}{|c|cc|ccc|c|}
				\hline
				\multirow{2}{*}{Decoders} & \multicolumn{2}{c|}{Arithmetic}                                   & \multicolumn{3}{c|}{LUTs}                                       & \multirow{2}{*}{Total} \\ \cline{2-6}
				& \multicolumn{1}{c|}{CN}                       & VN                       & \multicolumn{1}{c|}{$R_c=2/3$}  & \multicolumn{1}{c|}{$R_c=3/4$}  & $R_c=5/6$  &                        \\ \hline
				Proposed                  & \multicolumn{1}{c|}{\multirow{4}{*}{0.42 kB}} & \multirow{3}{*}{1.27 kB} & \multicolumn{3}{c|}{0.14 kB}                                          & 1.83 kB                \\ \cline{1-1} \cline{4-7} 
				Rate-specific             & \multicolumn{1}{c|}{}                         &                          & \multicolumn{1}{c|}{0.48 kB} & \multicolumn{1}{c|}{0.59 kB} & 0.59 kB & 3.35 kB                \\ \cline{1-1} \cline{4-7} 
				NMS \cite{Jinghu2005reduced}              & \multicolumn{1}{c|}{}                         &                          & \multicolumn{3}{c|}{-}                                                & 1.69 kB                \\ \cline{1-1} \cline{3-7}
				LUT-based \cite{Stark2020ib}        & \multicolumn{1}{c|}{}                        & -                        & \multicolumn{3}{c|}{35.63 kB}                                         & 36.05 kB               \\ \hline
			\end{tabular}
		}\vspace{-0.5cm}
	\end{table}
	\section{Conclusion}
	\label{section: conclusion}
	\vspace{-0.1cm}
	In this paper, we proposed a two-stage design approach to construct the rate-compatible MIM-QMS decoder, which significantly reduces the memory requirement for decoding.
	More specifically, the proposed design method first adopts the modified MIM-DE to construct the LUTs used for different code rates.
	The constructed LUTs are further optimized based on a merge function and their discrepancy values for two adjacent iterations.  
	We show that the proposed rate-compatible MIM-QMS decoder reduces the memory requirement for decoding by up to $94.92 \%$ compared to the rate-compatible LUT-based decoder with faster convergence speed in the moderate-to-high SNR region.
	In addition, there is only minor performance loss compared to both the rate-specific MIM-QMS decoders and the floating-pointing BP decoder.\vspace{-0.2cm}
	\bibliographystyle{ieeetr}

	\newpage
	 \appendix
		\subsection{The 4-bit MIM-QMS decoder for rate-$2/3$ 802.11n LDPC code}
		\begin{table}[h!]
			\Large
			\centering
			\setstretch{1.1}
			\caption{Reconstruction Functions $\phi_v$ and $\phi_{ch}$}
			\resizebox{0.5\textwidth}{!}{
			\begin{tabular}{l|cccccccccccccccc}
				\hline
				\multicolumn{1}{l|}{\multirow{2}{*}{Iteration}} & \multicolumn{16}{c}{$\phi_v(s)$, $s \in \mc{S}$} \\ \cline{2-17} 
				\multicolumn{1}{l|}{}  & 0  & 1  & 2  & 3   & 4    & 5   & 6   & 7 & 8 & 9 & 10 & 11 & 12 & 13 & 14 & 15 \\ \hline
				1	& 9  & 7  & 5  & 3  & 2  & 1  & 1  & 0  & 0  & -1  & -1  & -2  & -3  & -5  & -7  & -9  \\ 
				2	& 9  & 7  & 5  & 3  & 2  & 2  & 1  & 0  & 0  & -1  & -2  & -2  & -3  & -5  & -7  & -9  \\ 
				3	& 9  & 7  & 5  & 4  & 2  & 1  & 1  & 0  & 0  & -1  & -1  & -2  & -4  & -5  & -7  & -9  \\ 
				4	& 10  & 8  & 6  & 4  & 3  & 2  & 1  & 0  & 0  & -1  & -2  & -3  & -4  & -6  & -8  & -10  \\ 
				5	& 10  & 7  & 5  & 4  & 2  & 1  & 1  & 0  & 0  & -1  & -1  & -2  & -4  & -5  & -7  & -10  \\ 
				6	& 11  & 8  & 6  & 4  & 3  & 1  & 1  & 0  & 0  & -1  & -1  & -3  & -4  & -6  & -8  & -11  \\ 
				7	& 13  & 9  & 7  & 5  & 3  & 2  & 1  & 0  & 0  & -1  & -2  & -3  & -5  & -7  & -9  & -13  \\ 
				8	& 12  & 8  & 6  & 4  & 3  & 2  & 1  & 0  & 0  & -1  & -2  & -3  & -4  & -6  & -8  & -12  \\ 
				9	& 12  & 9  & 7  & 5  & 3  & 2  & 1  & 0  & 0  & -1  & -2  & -3  & -5  & -7  & -9  & -12  \\ 
				10	& 13  & 9  & 7  & 5  & 4  & 2  & 1  & 0  & 0  & -1  & -2  & -4  & -5  & -7  & -9  & -13  \\ 
				11	& 13  & 10  & 7  & 5  & 3  & 2  & 1  & 0  & 0  & -1  & -2  & -3  & -5  & -7  & -10  & -13  \\ 
				12	& 14  & 10  & 8  & 6  & 4  & 3  & 2  & 0  & 0  & -2  & -3  & -4  & -6  & -8  & -10  & -14  \\ 
				13	& 13  & 9  & 7  & 5  & 4  & 2  & 1  & 0  & 0  & -1  & -2  & -4  & -5  & -7  & -9  & -13  \\ 
				14	& 14  & 10  & 7  & 5  & 4  & 2  & 1  & 0  & 0  & -1  & -2  & -4  & -5  & -7  & -10  & -14  \\ 
				15	& 14  & 10  & 7  & 5  & 3  & 2  & 1  & 0  & 0  & -1  & -2  & -3  & -5  & -7  & -10  & -14  \\ 
				16	& 14  & 10  & 7  & 5  & 3  & 2  & 1  & 0  & 0  & -1  & -2  & -3  & -5  & -7  & -10  & -14  \\ 
				17	& 14  & 10  & 7  & 5  & 3  & 2  & 1  & 0  & 0  & -1  & -2  & -3  & -5  & -7  & -10  & -14  \\ 
				18	& 14  & 10  & 8  & 6  & 4  & 2  & 1  & 0  & 0  & -1  & -2  & -4  & -6  & -8  & -10  & -14  \\ 
				19	& 14  & 10  & 8  & 6  & 4  & 3  & 2  & 0  & 0  & -2  & -3  & -4  & -6  & -8  & -10  & -14  \\ 
				20	& 14  & 10  & 7  & 5  & 3  & 2  & 1  & 0  & 0  & -1  & -2  & -3  & -5  & -7  & -10  & -14  \\ 
				21	& 14  & 10  & 7  & 5  & 4  & 2  & 1  & 0  & 0  & -1  & -2  & -4  & -5  & -7  & -10  & -14  \\ 
				22	& 14  & 10  & 7  & 5  & 4  & 2  & 1  & 0  & 0  & -1  & -2  & -4  & -5  & -7  & -10  & -14  \\ 
				23	& 14  & 10  & 7  & 5  & 3  & 2  & 1  & 0  & 0  & -1  & -2  & -3  & -5  & -7  & -10  & -14  \\ 
				24	& 14  & 10  & 7  & 5  & 4  & 2  & 1  & 0  & 0  & -1  & -2  & -4  & -5  & -7  & -10  & -14  \\ 
				25	& 14  & 10  & 7  & 5  & 3  & 2  & 1  & 0  & 0  & -1  & -2  & -3  & -5  & -7  & -10  & -14  \\ 
				26	& 14  & 10  & 7  & 5  & 3  & 2  & 1  & 0  & 0  & -1  & -2  & -3  & -5  & -7  & -10  & -14  \\ 
				27	& 14  & 10  & 7  & 5  & 3  & 2  & 1  & 0  & 0  & -1  & -2  & -3  & -5  & -7  & -10  & -14  \\ 
				28	& 14  & 10  & 7  & 5  & 3  & 2  & 1  & 0  & 0  & -1  & -2  & -3  & -5  & -7  & -10  & -14  \\ 
				29	& 14  & 10  & 7  & 5  & 3  & 2  & 1  & 1  & -1  & -1  & -2  & -3  & -5  & -7  & -10  & -14  \\ 
				30	& 14  & 10  & 7  & 5  & 4  & 2  & 1  & 1  & -1  & -1  & -2  & -4  & -5  & -7  & -10  & -14  \\
				\hline
			\end{tabular}
		}
		\end{table}
		\begin{table}[!h]
			\Large
			\centering
			\setstretch{1.1}
			\resizebox{0.5\textwidth}{!}{
			\begin{tabular}{l|cccccccccccccccc}
				\hline
				\multicolumn{1}{l|}{\multirow{2}{*}{Iteration}} & \multicolumn{16}{c}{$\phi_{ch}(l)$, $l \in \mc{L}$} \\ \cline{2-17} 
				\multicolumn{1}{l|}{}  & 0  & 1  & 2  & 3   & 4    & 5   & 6   & 7 & 8 & 9 & 10 & 11 & 12 & 13 & 14 & 15 \\ \hline
				1-11	& 14  & 10  & 7  & 6  & 4  & 3  & 2  & 1  & -1  & -2  & -3  & -4  & -6  & -7  & -10  & -14  \\
				12	& 13  & 9  & 7  & 5  & 4  & 3  & 2  & 1  & -1  & -2  & -3  & -4  & -5  & -7  & -9  & -13  \\ 
				13	& 14  & 10  & 7  & 6  & 4  & 3  & 2  & 1  & -1  & -2  & -3  & -4  & -6  & -7  & -10  & -14  \\ 
				14	& 14  & 10  & 7  & 6  & 4  & 3  & 2  & 1  & -1  & -2  & -3  & -4  & -6  & -7  & -10  & -14  \\ 
				15	& 13  & 9  & 7  & 5  & 4  & 3  & 2  & 1  & -1  & -2  & -3  & -4  & -5  & -7  & -9  & -13  \\ 
				16	& 13  & 9  & 7  & 5  & 4  & 3  & 2  & 1  & -1  & -2  & -3  & -4  & -5  & -7  & -9  & -13  \\ 
				17	& 12  & 8  & 6  & 5  & 4  & 2  & 1  & 0  & 0  & -1  & -2  & -4  & -5  & -6  & -8  & -12  \\ 
				18	& 12  & 9  & 6  & 5  & 4  & 2  & 1  & 0  & 0  & -1  & -2  & -4  & -5  & -6  & -9  & -12  \\ 
				19	& 12  & 8  & 6  & 5  & 4  & 2  & 1  & 0  & 0  & -1  & -2  & -4  & -5  & -6  & -8  & -12  \\ 
				20	& 11  & 8  & 6  & 4  & 3  & 2  & 1  & 0  & 0  & -1  & -2  & -3  & -4  & -6  & -8  & -11  \\ 
				21	& 11  & 8  & 6  & 4  & 3  & 2  & 1  & 0  & 0  & -1  & -2  & -3  & -4  & -6  & -8  & -11  \\ 
				22	& 11  & 8  & 6  & 5  & 3  & 2  & 1  & 0  & 0  & -1  & -2  & -3  & -5  & -6  & -8  & -11  \\ 
				23	& 10  & 7  & 5  & 4  & 3  & 2  & 1  & 0  & 0  & -1  & -2  & -3  & -4  & -5  & -7  & -10  \\ 
				24	& 10  & 7  & 5  & 4  & 3  & 2  & 1  & 0  & 0  & -1  & -2  & -3  & -4  & -5  & -7  & -10  \\ 
				25	& 9  & 7  & 5  & 4  & 3  & 2  & 1  & 0  & 0  & -1  & -2  & -3  & -4  & -5  & -7  & -9  \\ 
				26	& 9  & 6  & 5  & 4  & 3  & 2  & 1  & 0  & 0  & -1  & -2  & -3  & -4  & -5  & -6  & -9  \\ 
				27	& 9  & 6  & 5  & 4  & 3  & 2  & 1  & 0  & 0  & -1  & -2  & -3  & -4  & -5  & -6  & -9  \\ 
				28	& 8  & 6  & 4  & 3  & 2  & 2  & 1  & 0  & 0  & -1  & -2  & -2  & -3  & -4  & -6  & -8  \\ 
				29	& 8  & 5  & 4  & 3  & 2  & 2  & 1  & 0  & 0  & -1  & -2  & -2  & -3  & -4  & -5  & -8  \\ 
				30	& 7  & 5  & 4  & 3  & 2  & 1  & 1  & 0  & 0  & -1  & -1  & -2  & -3  & -4  & -5  & -7  \\  \hline
			\end{tabular}
			}
		\end{table}
		\begin{table}[!h]
			\Large
			\centering
			\caption{Threshold sets $\Gamma_v$, $\Gamma_{ch}$, and $\Gamma_e$}
			\setstretch{1.1}
			\resizebox{0.5\textwidth}{!}{
			\begin{tabular}{l|ccccccccccccccc}
				\hline
				\multirow{2}{*}{Iteration} & \multicolumn{15}{c}{$\Gamma_v$}   \\ \cline{2-16} 
				& $\gamma_1$   & $\gamma_2$   & $\gamma_3$  & $\gamma_4$  & $\gamma_5$   & $\gamma_6$   & $\gamma_7$    & $\gamma_8$    & $\gamma_9$  & $\gamma_{10}$  & $\gamma_{11}$  & $\gamma_{12}$  & $\gamma_{13}$  & $\gamma_{14}$ & $\gamma_{15}$  \\ \hline
				1	& 12  & 9  & 7  & 5  & 4  & 3  & 2  & 1  & -1  & -2  & -3  & -4  & -6  & -8  & -11  \\ 
				2	& 12  & 9  & 7  & 5  & 4  & 3  & 2  & 1  & 0  & -1  & -3  & -4  & -6  & -8  & -11  \\ 
				3	& 13  & 10  & 8  & 6  & 4  & 3  & 2  & 0  & -1  & -2  & -3  & -5  & -7  & -9  & -12  \\ 
				4	& 13  & 9  & 7  & 5  & 4  & 3  & 2  & 1  & 0  & -1  & -2  & -4  & -6  & -8  & -12  \\ 
				5	& 13  & 10  & 8  & 6  & 4  & 3  & 2  & 1  & 0  & -1  & -2  & -4  & -6  & -8  & -12  \\ 
				6	& 15  & 11  & 8  & 6  & 4  & 3  & 2  & 0  & -1  & -2  & -3  & -5  & -7  & -10  & -14  \\ 
				7	& 14  & 10  & 7  & 5  & 4  & 2  & 1  & 0  & -1  & -2  & -3  & -4  & -6  & -9  & -13  \\ 
				8	& 16  & 12  & 9  & 7  & 5  & 3  & 2  & 1  & 0  & -1  & -2  & -4  & -6  & -9  & -13  \\ 
				9	& 16  & 12  & 9  & 7  & 5  & 4  & 2  & 1  & 0  & -1  & -3  & -4  & -6  & -9  & -13  \\ 
				10	& 15  & 11  & 8  & 6  & 4  & 3  & 2  & 0  & -1  & -2  & -3  & -5  & -7  & -10  & -14  \\ 
				11	& 16  & 12  & 9  & 7  & 5  & 4  & 2  & 1  & -1  & -3  & -4  & -6  & -8  & -11  & -15  \\ 
				12	& 15  & 11  & 9  & 7  & 5  & 3  & 2  & 1  & 0  & -1  & -2  & -4  & -6  & -8  & -12  \\ 
				13	& 15  & 11  & 8  & 6  & 4  & 3  & 2  & 0  & -1  & -2  & -3  & -5  & -7  & -10  & -14  \\ 
				14	& 16  & 11  & 8  & 6  & 4  & 3  & 2  & 0  & -1  & -2  & -3  & -5  & -7  & -10  & -15  \\ 
				15	& 14  & 10  & 7  & 5  & 4  & 3  & 2  & 1  & 0  & -1  & -2  & -4  & -6  & -9  & -13  \\ 
				16	& 16  & 11  & 8  & 6  & 4  & 2  & 1  & 0  & -1  & -2  & -3  & -5  & -7  & -10  & -15  \\ 
				17	& 15  & 11  & 8  & 6  & 4  & 3  & 2  & 1  & 0  & -1  & -2  & -4  & -6  & -8  & -12  \\ 
				18	& 16  & 12  & 9  & 7  & 5  & 4  & 2  & 1  & 0  & -1  & -3  & -4  & -6  & -9  & -13  \\ 
				19	& 16  & 11  & 8  & 5  & 3  & 2  & 1  & 0  & -1  & -2  & -3  & -5  & -7  & -10  & -15  \\ 
				20	& 13  & 9  & 6  & 4  & 3  & 2  & 1  & 0  & -1  & -2  & -3  & -5  & -7  & -9  & -13  \\ 
				21	& 13  & 9  & 7  & 5  & 3  & 2  & 1  & 0  & -1  & -2  & -3  & -4  & -6  & -8  & -12  \\ 
				22	& 15  & 10  & 7  & 5  & 3  & 2  & 1  & 0  & -1  & -2  & -3  & -5  & -7  & -10  & -15  \\ 
				23	& 13  & 9  & 7  & 5  & 4  & 3  & 2  & 1  & 0  & -1  & -2  & -4  & -6  & -8  & -12  \\ 
				24	& 14  & 10  & 7  & 5  & 4  & 3  & 2  & 1  & 0  & -1  & -2  & -4  & -6  & -9  & -13  \\ 
				25	& 13  & 9  & 7  & 5  & 3  & 2  & 1  & 0  & -1  & -2  & -3  & -4  & -6  & -8  & -12  \\ 
				26	& 13  & 9  & 7  & 5  & 3  & 2  & 1  & 0  & -1  & -2  & -3  & -4  & -6  & -8  & -12  \\ 
				27	& 13  & 9  & 7  & 5  & 4  & 3  & 2  & 1  & 0  & -1  & -2  & -4  & -6  & -8  & -12  \\ 
				28	& 13  & 9  & 6  & 4  & 3  & 2  & 1  & 0  & -1  & -2  & -3  & -4  & -6  & -8  & -12  \\ 
				29	& 14  & 10  & 8  & 6  & 4  & 3  & 2  & 1  & 0  & -1  & -2  & -3  & -5  & -8  & -12  \\ 
				30	& 12  & 8  & 6  & 5  & 4  & 3  & 2  & 1  & 0  & -1  & -2  & -3  & -5  & -7  & -11  \\ \hline
			\end{tabular}
			}
		\end{table}
		\begin{table}[h!]
			\centering
			\huge
			\setstretch{1.4}
			\resizebox{0.5\textwidth}{!}{
			\begin{tabular}{|c|c|c|c|c|c|c|c|c|c|c|c|c|c|c|}
				\hline
				\multicolumn{15}{|c|}{$\Gamma_{ch}$ (in LLR format)} \\ \hline
				$\gamma_1$   & $\gamma_2$   & $\gamma_3$  & $\gamma_4$  & $\gamma_5$   & $\gamma_6$   & $\gamma_7$    & $\gamma_8$    & $\gamma_9$  & $\gamma_{10}$  & $\gamma_{11}$  & $\gamma_{12}$  & $\gamma_{13}$  & $\gamma_{14}$ & $\gamma_{15}$  \\ \hline
				5.50  & 4.04  & 3.06  & 2.30  & 1.64  & 1.06  & 0.52  & 0  & -0.52  & -1.06  & -1.64  & -2.30  & -3.06  & -4.04  & -5.50  \\   \hline
			\end{tabular}
			}
		\end{table}
		\begin{table}[h!]
			\setstretch{1.1}
			\footnotesize
			\centering
			\begin{tabular}{l|c}
				\hline
				\multirow{2}{*}{Iteration} & $\Gamma_e$ \\ \cline{2-2} 
				& $\gamma_1$ \\ \hline
				1-5	& 0  \\ 
				6	& 1  \\ 
				7	& 0  \\ 
				8-28	& 1  \\
				29	& 0  \\ 
				30	& 1  \\ 
				 \hline
			\end{tabular}
		\end{table}
		\clearpage
		\newpage
		\subsection{The 4-bit MIM-QMS decoder for rate-$3/4$ 802.11n LDPC code}
		\begin{table}[h!]
			\Large
			\centering
			\setstretch{1.1}
			\caption{Reconstruction Functions $\phi_v$ and $\phi_{ch}$}
			\resizebox{0.5\textwidth}{!}{
				\begin{tabular}{l|cccccccccccccccc}
					\hline
					\multicolumn{1}{l|}{\multirow{2}{*}{Iteration}} & \multicolumn{16}{c}{$\phi_v(s)$, $s \in \mc{S}$} \\ \cline{2-17} 
					\multicolumn{1}{l|}{}  & 0  & 1  & 2  & 3   & 4    & 5   & 6   & 7 & 8 & 9 & 10 & 11 & 12 & 13 & 14 & 15 \\ \hline
					1	& 12  & 8  & 6  & 4  & 3  & 2  & 1  & 0  & 0  & -1  & -2  & -3  & -4  & -6  & -8  & -12  \\ 
					2	& 12  & 9  & 7  & 5  & 3  & 2  & 1  & 0  & 0  & -1  & -2  & -3  & -5  & -7  & -9  & -12  \\ 
					3	& 13  & 10  & 7  & 5  & 3  & 2  & 1  & 0  & 0  & -1  & -2  & -3  & -5  & -7  & -10  & -13  \\ 
					4-6	& 13  & 10  & 7  & 5  & 4  & 2  & 1  & 0  & 0  & -1  & -2  & -4  & -5  & -7  & -10  & -13  \\ 
					7	& 13  & 10  & 7  & 6  & 4  & 2  & 1  & 0  & 0  & -1  & -2  & -4  & -6  & -7  & -10  & -13  \\ 
					8	& 14  & 10  & 7  & 6  & 4  & 2  & 1  & 0  & 0  & -1  & -2  & -4  & -6  & -7  & -10  & -14  \\ 
					9	& 14  & 10  & 8  & 6  & 4  & 2  & 1  & 0  & 0  & -1  & -2  & -4  & -6  & -8  & -10  & -14  \\ 
					10	& 15  & 11  & 8  & 6  & 5  & 3  & 2  & 1  & -1  & -2  & -3  & -5  & -6  & -8  & -11  & -15  \\ 
					11	& 15  & 11  & 8  & 6  & 5  & 3  & 2  & 0  & 0  & -2  & -3  & -5  & -6  & -8  & -11  & -15  \\ 
					12	& 14  & 10  & 8  & 6  & 4  & 3  & 2  & 0  & 0  & -2  & -3  & -4  & -6  & -8  & -10  & -14  \\ 
					13	& 14  & 10  & 8  & 6  & 4  & 2  & 1  & 0  & 0  & -1  & -2  & -4  & -6  & -8  & -10  & -14  \\ 
					14	& 15  & 11  & 9  & 7  & 5  & 3  & 2  & 1  & -1  & -2  & -3  & -5  & -7  & -9  & -11  & -15  \\ 
					15	& 15  & 11  & 9  & 7  & 5  & 3  & 2  & 0  & 0  & -2  & -3  & -5  & -7  & -9  & -11  & -15  \\ 
					16	& 15  & 11  & 8  & 6  & 4  & 3  & 1  & 0  & 0  & -1  & -3  & -4  & -6  & -8  & -11  & -15  \\ 
					17	& 15  & 11  & 8  & 6  & 4  & 3  & 1  & 0  & 0  & -1  & -3  & -4  & -6  & -8  & -11  & -15  \\ 
					18	& 15  & 11  & 8  & 6  & 4  & 3  & 2  & 0  & 0  & -2  & -3  & -4  & -6  & -8  & -11  & -15  \\ 
					19	& 16  & 11  & 8  & 6  & 5  & 3  & 2  & 0  & 0  & -2  & -3  & -5  & -6  & -8  & -11  & -16  \\ 
					20	& 18  & 14  & 10  & 8  & 6  & 4  & 2  & 0  & 0  & -2  & -4  & -6  & -8  & -10  & -14  & -18  \\ 
					21	& 18  & 13  & 10  & 7  & 5  & 4  & 2  & 1  & -1  & -2  & -4  & -5  & -7  & -10  & -13  & -18  \\ 
					22	& 18  & 13  & 10  & 8  & 6  & 4  & 2  & 1  & -1  & -2  & -4  & -6  & -8  & -10  & -13  & -18  \\ 
					23	& 18  & 13  & 10  & 8  & 6  & 4  & 2  & 1  & -1  & -2  & -4  & -6  & -8  & -10  & -13  & -18  \\ 
					24	& 18  & 13  & 10  & 8  & 5  & 4  & 2  & 1  & -1  & -2  & -4  & -5  & -8  & -10  & -13  & -18  \\ 
					25	& 18  & 13  & 10  & 8  & 6  & 4  & 2  & 0  & 0  & -2  & -4  & -6  & -8  & -10  & -13  & -18  \\ 
					26	& 18  & 13  & 9  & 7  & 5  & 4  & 2  & 0  & 0  & -2  & -4  & -5  & -7  & -9  & -13  & -18  \\ 
					27	& 18  & 13  & 9  & 7  & 5  & 3  & 2  & 0  & 0  & -2  & -3  & -5  & -7  & -9  & -13  & -18  \\ 
					28	& 18  & 12  & 8  & 6  & 4  & 3  & 2  & 1  & -1  & -2  & -3  & -4  & -6  & -8  & -12  & -18  \\ 
					29	& 18  & 12  & 8  & 6  & 4  & 3  & 1  & 0  & 0  & -1  & -3  & -4  & -6  & -8  & -12  & -18  \\ 
					30	& 18  & 12  & 8  & 6  & 4  & 3  & 2  & 0  & 0  & -2  & -3  & -4  & -6  & -8  & -12  & -18  \\ 
					\hline
				\end{tabular}
			}
		\end{table}
		\begin{table}[!h]
			\Large
			\centering
			\setstretch{1.1}
			\resizebox{0.5\textwidth}{!}{
			\begin{tabular}{l|cccccccccccccccc}
					\hline
					\multicolumn{1}{l|}{\multirow{2}{*}{Iteration}} & \multicolumn{16}{c}{$\phi_{ch}(l)$, $l \in \mc{L}$} \\ \cline{2-17} 
					\multicolumn{1}{l|}{}  & 0  & 1  & 2  & 3   & 4    & 5   & 6   & 7 & 8 & 9 & 10 & 11 & 12 & 13 & 14 & 15 \\ \hline
				1-20	& 18  & 12  & 9  & 7  & 5  & 4  & 2  & 1  & -1  & -2  & -4  & -5  & -7  & -9  & -12  & -18  \\
				21	& 17  & 12  & 9  & 7  & 5  & 3  & 2  & 1  & -1  & -2  & -3  & -5  & -7  & -9  & -12  & -17  \\ 
				22	& 16  & 11  & 8  & 6  & 5  & 3  & 2  & 1  & -1  & -2  & -3  & -5  & -6  & -8  & -11  & -16  \\ 
				23	& 16  & 11  & 9  & 6  & 5  & 3  & 2  & 1  & -1  & -2  & -3  & -5  & -6  & -9  & -11  & -16  \\ 
				24	& 15  & 10  & 8  & 6  & 4  & 3  & 2  & 1  & -1  & -2  & -3  & -4  & -6  & -8  & -10  & -15  \\ 
				25	& 15  & 10  & 8  & 6  & 4  & 3  & 2  & 1  & -1  & -2  & -3  & -4  & -6  & -8  & -10  & -15  \\ 
				26	& 13  & 9  & 7  & 5  & 4  & 3  & 2  & 1  & -1  & -2  & -3  & -4  & -5  & -7  & -9  & -13  \\ 
				27	& 12  & 8  & 6  & 5  & 4  & 2  & 1  & 0  & 0  & -1  & -2  & -4  & -5  & -6  & -8  & -12  \\ 
				28	& 12  & 8  & 6  & 5  & 3  & 2  & 1  & 0  & 0  & -1  & -2  & -3  & -5  & -6  & -8  & -12  \\ 
				29	& 11  & 8  & 6  & 4  & 3  & 2  & 1  & 0  & 0  & -1  & -2  & -3  & -4  & -6  & -8  & -11  \\ 
				30	& 10  & 7  & 5  & 4  & 3  & 2  & 1  & 0  & 0  & -1  & -2  & -3  & -4  & -5  & -7  & -10  \\ 
				\hline
			\end{tabular}
			}
		\end{table}
		\begin{table}[!h]
			\Large
			\centering
			\caption{Threshold sets $\Gamma_v$, $\Gamma_{ch}$, and $\Gamma_e$}
			\setstretch{1.1}
			\resizebox{0.5\textwidth}{!}{
				\begin{tabular}{l|ccccccccccccccc}
					\hline
					\multirow{2}{*}{Iteration} & \multicolumn{15}{c}{$\Gamma_v$}   \\ \cline{2-16} 
					& $\gamma_1$   & $\gamma_2$   & $\gamma_3$  & $\gamma_4$  & $\gamma_5$   & $\gamma_6$   & $\gamma_7$    & $\gamma_8$    & $\gamma_9$  & $\gamma_{10}$  & $\gamma_{11}$  & $\gamma_{12}$  & $\gamma_{13}$  & $\gamma_{14}$ & $\gamma_{15}$  \\ \hline
					1	& 15  & 11  & 9  & 7  & 5  & 3  & 2  & 0  & -1  & -2  & -4  & -6  & -8  & -10  & -14  \\ 
					2	& 16  & 12  & 9  & 7  & 5  & 3  & 2  & 0  & -1  & -2  & -4  & -6  & -8  & -11  & -15  \\ 
					3	& 16  & 12  & 9  & 7  & 5  & 3  & 2  & 1  & -1  & -2  & -4  & -6  & -8  & -11  & -15  \\ 
					4	& 16  & 12  & 9  & 7  & 5  & 3  & 1  & 0  & -1  & -2  & -4  & -6  & -8  & -11  & -15  \\ 
					5	& 16  & 12  & 9  & 7  & 5  & 3  & 2  & 1  & 0  & -2  & -4  & -6  & -8  & -11  & -15  \\ 
					6	& 16  & 12  & 9  & 7  & 5  & 3  & 2  & 1  & 0  & -2  & -4  & -6  & -8  & -11  & -15  \\ 
					7	& 17  & 12  & 9  & 7  & 5  & 3  & 2  & 1  & 0  & -2  & -4  & -6  & -8  & -11  & -16  \\ 
					8	& 17  & 12  & 9  & 7  & 5  & 3  & 2  & 1  & 0  & -2  & -4  & -6  & -8  & -11  & -16  \\ 
					9	& 17  & 12  & 9  & 7  & 5  & 3  & 1  & 0  & -2  & -4  & -6  & -8  & -10  & -13  & -18  \\ 
					10	& 18  & 13  & 10  & 8  & 6  & 4  & 2  & 1  & -1  & -3  & -5  & -7  & -9  & -12  & -17  \\ 
					11	& 17  & 12  & 9  & 7  & 5  & 4  & 2  & 1  & -1  & -3  & -4  & -6  & -8  & -11  & -16  \\ 
					12	& 17  & 12  & 9  & 7  & 5  & 3  & 2  & 0  & -1  & -2  & -4  & -6  & -8  & -11  & -16  \\ 
					13	& 17  & 12  & 9  & 7  & 5  & 3  & 1  & 0  & -2  & -4  & -6  & -8  & -10  & -13  & -18  \\ 
					14	& 18  & 13  & 10  & 8  & 6  & 4  & 2  & 0  & -1  & -3  & -5  & -7  & -9  & -12  & -17  \\ 
					15-18	& 17  & 12  & 9  & 7  & 5  & 3  & 2  & 0  & -1  & -2  & -4  & -6  & -8  & -11  & -16  \\
					19	& 20  & 15  & 11  & 8  & 6  & 4  & 2  & 1  & -1  & -3  & -5  & -7  & -10  & -14  & -19  \\ 
					20	& 20  & 14  & 10  & 7  & 5  & 3  & 1  & 0  & -2  & -4  & -6  & -8  & -10  & -13  & -19  \\ 
					21	& 20  & 14  & 10  & 8  & 6  & 4  & 2  & 0  & -2  & -4  & -6  & -8  & -11  & -14  & -20  \\ 
					22	& 18  & 13  & 10  & 8  & 6  & 4  & 2  & 0  & -1  & -3  & -5  & -7  & -9  & -12  & -17  \\ 
					23	& 20  & 14  & 11  & 8  & 6  & 4  & 2  & 0  & -1  & -3  & -5  & -7  & -10  & -13  & -19  \\ 
					24	& 17  & 12  & 9  & 7  & 5  & 3  & 1  & 0  & -1  & -3  & -5  & -7  & -10  & -13  & -18  \\ 
					25	& 19  & 13  & 10  & 8  & 6  & 4  & 2  & 1  & -1  & -3  & -5  & -7  & -9  & -12  & -18  \\ 
					26	& 18  & 12  & 9  & 7  & 5  & 3  & 2  & 1  & 0  & -2  & -4  & -6  & -8  & -11  & -17  \\ 
					27	& 17  & 11  & 8  & 6  & 4  & 3  & 2  & 1  & -1  & -2  & -3  & -5  & -7  & -10  & -16  \\ 
					28	& 18  & 12  & 9  & 7  & 5  & 3  & 2  & 1  & 0  & -1  & -2  & -4  & -6  & -9  & -15  \\ 
					29	& 16  & 11  & 8  & 6  & 4  & 3  & 1  & 0  & -1  & -2  & -3  & -5  & -7  & -10  & -15  \\ 
					30	& 16  & 11  & 8  & 6  & 4  & 3  & 2  & 1  & 0  & -1  & -2  & -4  & -6  & -9  & -14  \\
					\hline
				\end{tabular}
			}
		\end{table}
		\begin{table}[h!]
			\centering
			\huge
			\setstretch{1.4}
			\resizebox{0.5\textwidth}{!}{
				\begin{tabular}{|c|c|c|c|c|c|c|c|c|c|c|c|c|c|c|}
					\hline
					\multicolumn{15}{|c|}{$\Gamma_{ch}$ (in LLR format)} \\ \hline
					$\gamma_1$   & $\gamma_2$   & $\gamma_3$  & $\gamma_4$  & $\gamma_5$   & $\gamma_6$   & $\gamma_7$    & $\gamma_8$    & $\gamma_9$  & $\gamma_{10}$  & $\gamma_{11}$  & $\gamma_{12}$  & $\gamma_{13}$  & $\gamma_{14}$ & $\gamma_{15}$  \\ \hline
					6.02  & 4.38  & 3.32  & 2.48  & 1.78  & 1.14  & 0.56  & 0  & -0.56  & -1.14  & -1.78  & -2.48  & -3.32  & -4.38  & -6.02  \\   \hline
				\end{tabular}
			}
		\end{table}
		\begin{table}[h!]
			\setstretch{1.1}
			\footnotesize
			\centering
			\begin{tabular}{l|c}
				\hline
				\multirow{2}{*}{Iteration} & $\Gamma_e$ \\ \cline{2-2} 
				& $\gamma_1$ \\ \hline
				1-11	& 1  \\ 
				12	& 0  \\ 
				13	& 1  \\ 
				14	& 0  \\ 
				15	& 1  \\ 
				16	& 1  \\ 
				17-28	& 0  \\ 
				29	& 1  \\ 
				30	& 1  \\ 
				\hline
			\end{tabular}
		\end{table}
		\clearpage
		\newpage
		\subsection{The 4-bit MIM-QMS decoder for rate-$5/6$ 802.11n LDPC code}
		\begin{table}[h!]
			\Large
			\centering
			\setstretch{1.1}
			\caption{Reconstruction Functions $\phi_v$ and $\phi_{ch}$}
			\resizebox{0.5\textwidth}{!}{
				\begin{tabular}{l|cccccccccccccccc}
					\hline
					\multicolumn{1}{l|}{\multirow{2}{*}{Iteration}} & \multicolumn{16}{c}{$\phi_v(s)$, $s \in \mc{S}$} \\ \cline{2-17} 
					\multicolumn{1}{l|}{}  & 0  & 1  & 2  & 3   & 4    & 5   & 6   & 7 & 8 & 9 & 10 & 11 & 12 & 13 & 14 & 15 \\ \hline
					1	& 16  & 11  & 8  & 6  & 4  & 2  & 1  & 0  & 0  & -1  & -2  & -4  & -6  & -8  & -11  & -16  \\ 
					2	& 16  & 11  & 9  & 6  & 4  & 3  & 1  & 0  & 0  & -1  & -3  & -4  & -6  & -9  & -11  & -16  \\ 
					3-6	& 17  & 12  & 9  & 7  & 5  & 3  & 2  & 0  & 0  & -2  & -3  & -5  & -7  & -9  & -12  & -17  \\ 
					7	& 17  & 13  & 10  & 7  & 5  & 3  & 2  & 0  & 0  & -2  & -3  & -5  & -7  & -10  & -13  & -17  \\ 
					8	& 18  & 13  & 10  & 7  & 5  & 3  & 2  & 0  & 0  & -2  & -3  & -5  & -7  & -10  & -13  & -18  \\ 
					9	& 18  & 13  & 10  & 7  & 5  & 3  & 1  & 0  & 0  & -1  & -3  & -5  & -7  & -10  & -13  & -18  \\ 
					10-13	& 18  & 13  & 10  & 7  & 5  & 3  & 2  & 1  & -1  & -2  & -3  & -5  & -7  & -10  & -13  & -18  \\
					14	& 18  & 14  & 11  & 8  & 5  & 3  & 2  & 1  & -1  & -2  & -3  & -5  & -8  & -11  & -14  & -18  \\ 
					15	& 18  & 14  & 10  & 8  & 5  & 3  & 2  & 1  & -1  & -2  & -3  & -5  & -8  & -10  & -14  & -18  \\ 
					16	& 19  & 14  & 11  & 8  & 6  & 4  & 2  & 1  & -1  & -2  & -4  & -6  & -8  & -11  & -14  & -19  \\ 
					17	& 19  & 14  & 11  & 8  & 5  & 4  & 2  & 1  & -1  & -2  & -4  & -5  & -8  & -11  & -14  & -19  \\ 
					18	& 18  & 14  & 10  & 8  & 5  & 4  & 2  & 1  & -1  & -2  & -4  & -5  & -8  & -10  & -14  & -18  \\ 
					19	& 19  & 14  & 10  & 8  & 5  & 4  & 2  & 1  & -1  & -2  & -4  & -5  & -8  & -10  & -14  & -19  \\ 
					20	& 20  & 14  & 11  & 8  & 6  & 4  & 2  & 1  & -1  & -2  & -4  & -6  & -8  & -11  & -14  & -20  \\ 
					21	& 20  & 15  & 11  & 8  & 6  & 4  & 2  & 1  & -1  & -2  & -4  & -6  & -8  & -11  & -15  & -20  \\ 
					22	& 21  & 15  & 11  & 8  & 6  & 4  & 2  & 1  & -1  & -2  & -4  & -6  & -8  & -11  & -15  & -21  \\ 
					23	& 21  & 16  & 12  & 9  & 7  & 4  & 3  & 1  & -1  & -3  & -4  & -7  & -9  & -12  & -16  & -21  \\ 
					24	& 22  & 17  & 12  & 9  & 7  & 4  & 2  & 1  & -1  & -2  & -4  & -7  & -9  & -12  & -17  & -22  \\ 
					25	& 23  & 17  & 13  & 10  & 7  & 4  & 3  & 1  & -1  & -3  & -4  & -7  & -10  & -13  & -17  & -23  \\ 
					26	& 25  & 19  & 14  & 10  & 7  & 5  & 3  & 1  & -1  & -3  & -5  & -7  & -10  & -14  & -19  & -25  \\ 
					27	& 25  & 19  & 14  & 11  & 8  & 5  & 3  & 1  & -1  & -3  & -5  & -8  & -11  & -14  & -19  & -25  \\ 
					28	& 25  & 19  & 14  & 10  & 8  & 5  & 3  & 1  & -1  & -3  & -5  & -8  & -10  & -14  & -19  & -25  \\ 
					29	& 25  & 18  & 12  & 8  & 6  & 4  & 2  & 1  & -1  & -2  & -4  & -6  & -8  & -12  & -18  & -25  \\ 
					30	& 25  & 17  & 13  & 10  & 7  & 5  & 3  & 1  & -1  & -3  & -5  & -7  & -10  & -13  & -17  & -25  \\ 
					\hline
				\end{tabular}
			}
		\end{table}
		\begin{table}[!h]
			\Large
			\centering
			\setstretch{1.1}
			\resizebox{0.5\textwidth}{!}{
				\begin{tabular}{l|cccccccccccccccc}
					\hline
					\multicolumn{1}{l|}{\multirow{2}{*}{Iteration}} & \multicolumn{16}{c}{$\phi_{ch}(l)$, $l \in \mc{L}$} \\ \cline{2-17} 
					\multicolumn{1}{l|}{}  & 0  & 1  & 2  & 3   & 4    & 5   & 6   & 7 & 8 & 9 & 10 & 11 & 12 & 13 & 14 & 15 \\ \hline
					1-25	& 25  & 17  & 13  & 10  & 7  & 5  & 3  & 1  & -1  & -3  & -5  & -7  & -10  & -13  & -17  & -25  \\ 
					26	& 25  & 17  & 13  & 9  & 7  & 5  & 3  & 1  & -1  & -3  & -5  & -7  & -9  & -13  & -17  & -25  \\ 
					27	& 23  & 16  & 12  & 9  & 6  & 4  & 3  & 1  & -1  & -3  & -4  & -6  & -9  & -12  & -16  & -23  \\ 
					28	& 20  & 14  & 10  & 8  & 6  & 4  & 2  & 1  & -1  & -2  & -4  & -6  & -8  & -10  & -14  & -20  \\ 
					29	& 18  & 12  & 9  & 7  & 5  & 3  & 2  & 1  & -1  & -2  & -3  & -5  & -7  & -9  & -12  & -18  \\ 
					30	& 16  & 11  & 8  & 6  & 4  & 3  & 2  & 1  & -1  & -2  & -3  & -4  & -6  & -8  & -11  & -16  \\ 
					\hline
				\end{tabular}
			}
		\end{table}
		\begin{table}[!h]
			\Large
			\centering
			\caption{Threshold sets $\Gamma_v$, $\Gamma_{ch}$, and $\Gamma_e$}
			\setstretch{1.1}
			\resizebox{0.5\textwidth}{!}{
				\begin{tabular}{l|ccccccccccccccc}
					\hline
					\multirow{2}{*}{Iteration} & \multicolumn{15}{c}{$\Gamma_v$}   \\ \cline{2-16} 
					& $\gamma_1$   & $\gamma_2$   & $\gamma_3$  & $\gamma_4$  & $\gamma_5$   & $\gamma_6$   & $\gamma_7$    & $\gamma_8$    & $\gamma_9$  & $\gamma_{10}$  & $\gamma_{11}$  & $\gamma_{12}$  & $\gamma_{13}$  & $\gamma_{14}$ & $\gamma_{15}$  \\ \hline
					1	& 20  & 15  & 12  & 9  & 6  & 4  & 2  & 0  & -2  & -4  & -6  & -8  & -11  & -14  & -19  \\ 
					2	& 22  & 17  & 13  & 10  & 7  & 5  & 3  & 1  & -1  & -3  & -5  & -7  & -10  & -14  & -20  \\ 
					3	& 21  & 15  & 11  & 8  & 6  & 4  & 2  & 0  & -2  & -4  & -6  & -9  & -12  & -16  & -21  \\ 
					4	& 23  & 17  & 13  & 10  & 7  & 5  & 3  & 1  & -1  & -3  & -5  & -7  & -10  & -14  & -20  \\ 
					5	& 21  & 15  & 11  & 8  & 6  & 4  & 2  & 0  & -2  & -4  & -6  & -9  & -12  & -16  & -22  \\ 
					6	& 23  & 17  & 13  & 10  & 7  & 5  & 3  & 1  & -1  & -3  & -5  & -7  & -10  & -14  & -20  \\ 
					7	& 22  & 16  & 12  & 9  & 6  & 4  & 2  & 0  & -2  & -4  & -6  & -9  & -12  & -16  & -22  \\ 
					8	& 22  & 16  & 12  & 9  & 6  & 4  & 2  & 1  & -1  & -3  & -5  & -8  & -11  & -15  & -21  \\ 
					9	& 23  & 17  & 13  & 10  & 7  & 5  & 3  & 1  & -1  & -3  & -5  & -7  & -10  & -14  & -20  \\ 
					10	& 22  & 16  & 12  & 9  & 6  & 4  & 2  & 0  & -2  & -4  & -6  & -8  & -11  & -15  & -21  \\ 
					11	& 22  & 16  & 12  & 9  & 7  & 5  & 3  & 1  & -1  & -3  & -5  & -8  & -11  & -15  & -21  \\ 
					12	& 22  & 16  & 12  & 9  & 6  & 4  & 2  & 0  & -2  & -4  & -6  & -8  & -11  & -15  & -21  \\ 
					13	& 23  & 17  & 13  & 10  & 7  & 5  & 3  & 1  & -1  & -3  & -5  & -8  & -11  & -15  & -21  \\ 
					14	& 22  & 16  & 12  & 9  & 7  & 5  & 3  & 1  & -1  & -3  & -5  & -8  & -11  & -15  & -21  \\ 
					15	& 23  & 17  & 13  & 10  & 7  & 5  & 3  & 1  & -1  & -3  & -5  & -8  & -11  & -15  & -21  \\ 
					16	& 23  & 17  & 13  & 10  & 7  & 5  & 3  & 1  & -1  & -3  & -5  & -8  & -11  & -15  & -22  \\ 
					17	& 22  & 16  & 12  & 9  & 7  & 5  & 3  & 1  & -1  & -3  & -5  & -8  & -11  & -15  & -21  \\ 
					18	& 23  & 16  & 12  & 9  & 7  & 5  & 3  & 1  & -1  & -3  & -5  & -8  & -11  & -15  & -22  \\ 
					19	& 24  & 17  & 13  & 10  & 7  & 5  & 3  & 1  & -1  & -3  & -5  & -8  & -11  & -15  & -22  \\ 
					20	& 24  & 17  & 13  & 10  & 7  & 5  & 3  & 1  & -1  & -3  & -5  & -8  & -11  & -15  & -22  \\ 
					21	& 24  & 17  & 12  & 9  & 6  & 4  & 2  & 0  & -2  & -4  & -6  & -9  & -12  & -16  & -23  \\ 
					22	& 24  & 17  & 13  & 10  & 7  & 5  & 3  & 1  & -2  & -4  & -6  & -9  & -12  & -16  & -23  \\ 
					23	& 25  & 18  & 13  & 10  & 7  & 4  & 2  & 0  & -2  & -4  & -6  & -9  & -12  & -17  & -24  \\ 
					24	& 25  & 18  & 13  & 10  & 7  & 4  & 2  & 0  & -2  & -4  & -6  & -9  & -12  & -17  & -24  \\ 
					25	& 27  & 19  & 14  & 10  & 7  & 5  & 3  & 1  & -1  & -3  & -6  & -9  & -13  & -18  & -26  \\ 
					26	& 29  & 21  & 16  & 12  & 9  & 6  & 4  & 2  & -1  & -3  & -6  & -9  & -13  & -18  & -26  \\ 
					27	& 28  & 20  & 15  & 11  & 8  & 5  & 3  & 0  & -2  & -4  & -7  & -10  & -14  & -19  & -27  \\ 
					28	& 25  & 17  & 12  & 8  & 6  & 4  & 2  & 1  & -1  & -3  & -5  & -7  & -11  & -16  & -24  \\ 
					29	& 24  & 17  & 13  & 10  & 7  & 5  & 3  & 1  & -1  & -4  & -6  & -9  & -12  & -16  & -23  \\ 
					30	& 25  & 17  & 13  & 10  & 8  & 5  & 3  & 1  & 0  & -2  & -4  & -7  & -10  & -14  & -22  \\
					\hline
				\end{tabular}
			}
		\end{table}
		\begin{table}[h!]
			\centering
			\huge
			\setstretch{1.4}
			\resizebox{0.5\textwidth}{!}{
				\begin{tabular}{|c|c|c|c|c|c|c|c|c|c|c|c|c|c|c|}
					\hline
					\multicolumn{15}{|c|}{$\Gamma_{ch}$ (in LLR format)} \\ \hline
					$\gamma_1$   & $\gamma_2$   & $\gamma_3$  & $\gamma_4$  & $\gamma_5$   & $\gamma_6$   & $\gamma_7$    & $\gamma_8$    & $\gamma_9$  & $\gamma_{10}$  & $\gamma_{11}$  & $\gamma_{12}$  & $\gamma_{13}$  & $\gamma_{14}$ & $\gamma_{15}$  \\ \hline
					6.64  & 4.80  & 3.60  & 2.68  & 1.90  & 1.24  & 0.60  & 0  & -0.60  & -1.24  & -1.90  & -2.68 & -3.60  & -4.80  & -6.64  \\   \hline
				\end{tabular}
			}
		\end{table}
		\begin{table}[h!]
			\setstretch{1.1}
			\footnotesize
			\centering
			\begin{tabular}{l|c}
				\hline
				\multirow{2}{*}{Iteration} & $\Gamma_e$ \\ \cline{2-2} 
				& $\gamma_1$ \\ \hline
				1-4	& 0  \\ 
				5-7	& 1  \\ 
				8	& 0  \\ 
				9-29	& 1  \\
				30	& 0  \\ 
				\hline
			\end{tabular}
		\end{table}
		\clearpage
		\newpage
		\subsection{The 4-bit rate-compatible MIM-QMS decoder for 802.11n LDPC codes}
		\begin{table}[h!]
			\Large
			\centering
			\setstretch{1.1}
			\caption{Reconstruction Functions $\phi_v$ and $\phi_{ch}$}
			\resizebox{0.5\textwidth}{!}{
				\begin{tabular}{l|cccccccccccccccc}
					\hline
					\multicolumn{1}{l|}{\multirow{2}{*}{Iteration}} & \multicolumn{16}{c}{$\phi_v(s)$, $s \in \mc{S}$} \\ \cline{2-17} 
					\multicolumn{1}{l|}{}  & 0  & 1  & 2  & 3   & 4    & 5   & 6   & 7 & 8 & 9 & 10 & 11 & 12 & 13 & 14 & 15 \\ \hline
					1-14	& 11  & 8  & 6  & 4  & 3  & 2  & 1  & 0  & 0  & -1  & -2  & -3  & -4  & -6  & -8  & -11  \\ 
					15	& 13  & 10  & 7  & 5  & 3  & 2  & 1  & 0  & 0  & -1  & -2  & -3  & -5  & -7  & -10  & -13  \\ 
					16	& 13  & 10  & 7  & 5  & 3  & 2  & 1  & 0  & 0  & -1  & -2  & -3  & -5  & -7  & -10  & -13  \\ 
					17-30	& 14  & 10  & 7  & 5  & 3  & 2  & 1  & 0  & 0  & -1  & -2  & -3  & -5  & -7  & -10  & -14  \\
					\hline
				\end{tabular}
			}
		\end{table}
		\begin{table}[!h]
			\Large
			\centering
			\setstretch{1.1}
			\resizebox{0.5\textwidth}{!}{
				\begin{tabular}{l|cccccccccccccccc}
					\hline
					\multicolumn{1}{l|}{\multirow{2}{*}{Iteration}} & \multicolumn{16}{c}{$\phi_{ch}(l)$, $l \in \mc{L}$} \\ \cline{2-17} 
					\multicolumn{1}{l|}{}  & 0  & 1  & 2  & 3   & 4    & 5   & 6   & 7 & 8 & 9 & 10 & 11 & 12 & 13 & 14 & 15 \\ \hline
					1-21	& 14  & 10  & 7  & 5  & 4  & 3  & 2  & 1  & -1  & -2  & -3  & -4  & -5  & -7  & -10  & -14  \\ 
					22-30	& 10  & 7  & 5  & 4  & 3  & 2  & 1  & 0  & 0  & -1  & -2  & -3  & -4  & -5  & -7  & -10  \\ 
					\hline
				\end{tabular}
			}
		\end{table}
		\begin{table}[!h]
			\Large
			\centering
			\caption{Threshold sets $\Gamma_v$, $\Gamma_{ch}$, and $\Gamma_e$}
			\setstretch{1.1}
			\resizebox{0.5\textwidth}{!}{
				\begin{tabular}{l|ccccccccccccccc}
					\hline
					\multirow{2}{*}{Iteration} & \multicolumn{15}{c}{$\Gamma_v$}   \\ \cline{2-16} 
					& $\gamma_1$   & $\gamma_2$   & $\gamma_3$  & $\gamma_4$  & $\gamma_5$   & $\gamma_6$   & $\gamma_7$    & $\gamma_8$    & $\gamma_9$  & $\gamma_{10}$  & $\gamma_{11}$  & $\gamma_{12}$  & $\gamma_{13}$  & $\gamma_{14}$ & $\gamma_{15}$  \\ \hline
					1-3	& 12  & 9  & 7  & 5  & 3  & 2  & 1  & 0  & -1  & -2  & -3  & -4  & -6  & -8  & -11  \\ 
					4-11	& 13  & 9  & 7  & 5  & 4  & 3  & 2  & 1  & 0  & -1  & -2  & -4  & -6  & -8  & -12  \\ 
					12	& 14  & 10  & 7  & 5  & 4  & 3  & 2  & 1  & 0  & -1  & -2  & -4  & -6  & -9  & -13  \\ 
					13	& 14  & 10  & 7  & 5  & 4  & 3  & 2  & 1  & 0  & -1  & -2  & -4  & -6  & -9  & -13  \\ 
					14-18	& 15  & 11  & 8  & 6  & 4  & 3  & 2  & 1  & 0  & -1  & -2  & -4  & -6  & -9  & -13  \\ 
					19	& 15  & 10  & 7  & 5  & 3  & 2  & 1  & 0  & -1  & -2  & -3  & -5  & -7  & -10  & -14  \\ 
					20	& 15  & 10  & 7  & 5  & 3  & 2  & 1  & 0  & -1  & -2  & -3  & -5  & -7  & -10  & -14  \\ 
					21	& 16  & 12  & 9  & 7  & 5  & 3  & 2  & 1  & 0  & -1  & -2  & -4  & -6  & -9  & -13  \\ 
					22	& 15  & 11  & 8  & 6  & 4  & 3  & 2  & 1  & 0  & -1  & -3  & -4  & -6  & -9  & -13  \\ 
					23	& 15  & 11  & 8  & 6  & 4  & 3  & 2  & 1  & 0  & -1  & -3  & -4  & -6  & -9  & -13  \\ 
					24-27	& 14  & 9  & 7  & 5  & 3  & 2  & 1  & 0  & -1  & -2  & -3  & -4  & -6  & -8  & -13  \\
					28	& 14  & 10  & 8  & 6  & 4  & 3  & 2  & 1  & 0  & -1  & -2  & -3  & -5  & -7  & -11  \\ 
					29	& 14  & 10  & 8  & 6  & 4  & 3  & 2  & 1  & 0  & -1  & -2  & -3  & -5  & -7  & -11  \\ 
					30	& 12  & 8  & 6  & 5  & 4  & 3  & 2  & 1  & 0  & -1  & -2  & -3  & -5  & -7  & -11  \\ 
					\hline
				\end{tabular}
			}
		\end{table}
		\begin{table}[h!]
			\centering
			\huge
			\setstretch{1.4}
			\resizebox{0.5\textwidth}{!}{
				\begin{tabular}{|c|c|c|c|c|c|c|c|c|c|c|c|c|c|c|}
					\hline
					\multicolumn{15}{|c|}{$\Gamma_{ch}$ (in LLR format)} \\ \hline
					$\gamma_1$   & $\gamma_2$   & $\gamma_3$  & $\gamma_4$  & $\gamma_5$   & $\gamma_6$   & $\gamma_7$    & $\gamma_8$    & $\gamma_9$  & $\gamma_{10}$  & $\gamma_{11}$  & $\gamma_{12}$  & $\gamma_{13}$  & $\gamma_{14}$ & $\gamma_{15}$  \\ \hline
					6.06  & 4.42  & 3.34  & 2.50  & 1.78 & 1.16  & 0.56  & 0  & -0.56  & -1.16  & -1.78  & -2.50  & -3.34  & -4.42  & -6.06  \\   \hline
				\end{tabular}
			}
		\end{table}
		\begin{table}[h!]
			\setstretch{1.1}
			\footnotesize
			\centering
			\begin{tabular}{l|c}
				\hline
				\multirow{2}{*}{Iteration} & $\Gamma_e$ \\ \cline{2-2} 
				& $\gamma_1$ \\ \hline
				1-30	& 1  \\ 
				\hline
			\end{tabular}
		\end{table}
\end{document}